\documentclass[preprint2]{aastex631}

\usepackage{amsmath}
\usepackage{xspace}
\usepackage{xcolor}
\usepackage{graphicx}
\usepackage{epstopdf}
\usepackage{placeins}
\usepackage{multirow}

\newcommand{\degree}{\mbox{$^\circ$}}
\newcommand{\chxoh}{\mbox{CH$_3$OH}}
\newcommand{\funit}{\mbox{mJy~beam$^{-1}$}}
\newcommand{\funitv}{\mbox{mJy~beam$^{-1}$~km~s$^{-1}$}}
\newcommand{\kms}{\mbox{km~s$^{-1}$}}

\newcommand{\chxohl}{\mbox{CH$_3$OH\,7$_{\rm k}$--6$_{\rm k}$}}

\shorttitle{HOPS\,373}
\shortauthors{Lee et al.}
\graphicspath{{./}{figures/}}

\begin{document}

\title{Multiple Jets in the bursting protostar HOPS\,373SW}

\author[0000-0002-0226-9295]{Seokho Lee}
\affiliation{Korea Astronomy and Space Science Institute 776 Daedeok-daero, Yuseong-gu, Daejeon 34055, Republic of Korea}
\email{seokholee@kasi.re.kr}

\author[0000-0003-3119-2087]{Jeong-Eun Lee}
\affiliation{Department of Physics and Astronomy, Seoul National University, 1 Gwanak-ro, Gwanak-gu, Seoul 08826, Korea}
\affil{SNU Astronomy Research Center, Seoul National University, 1 Gwanak-ro, Gwanak-gu, Seoul 08826, Republic of Korea}
\email{lee.jeongeun@snu.ac.kr}

\author[0000-0002-6773-459X]{Doug Johnstone}
\affiliation{NRC Herzberg Astronomy and Astrophysics, 5071 West Saanich Rd, Victoria, BC, V9E 2E7, Canada }
\affiliation{Department of Physics and Astronomy, University of Victoria, Victoria, BC, V8P 5C2, Canada}

\author[0000-0002-7154-6065]{Gregory J. Herczeg}
\affiliation{Kavli Institute for Astronomy and Astrophysics, Peking University, Yiheyuan Lu 5, Haidian Qu, 100871, Beijing, PR China}

\author[0000-0003-3283-6884]{Yuri Aikawa}
\affiliation{Department of Astronomy, University of Tokyo, 7-3-1 Hongo, Bunkyo-ku, Tokyo 113-0033, Japan}




\begin{abstract}
We present the outflows detected in HOPS\,373SW, a protostar undergoing a modest $30\%$ brightness increase at 850 $\mu$m. 
Atacama Large Millimeter/submillimeter Array (ALMA) observations of shock tracers, including SiO\,8--7, \chxohl, and $^{12}$CO\,3--2 emission, reveal several outflow features around HOPS\,373SW. 
The knots in the extremely high-velocity SiO emission reveal the wiggle of the jet, for which a simple model derives a 37\degree\ inclination angle of the jet to the plane of the sky, a jet velocity of 90~\kms, and a period of 50 years. 
The slow SiO and \chxoh\ emission traces U-shaped bow shocks surrounding the two CO outflows. One outflow is associated with the high-velocity jets, while the other is observed  to be close to the plane of the sky. The misaligned outflows imply that previous episodic accretion events have either reoriented HOPS\,373SW or that it is an unresolved protostellar binary system with misaligned outflows. 
\end{abstract}



\section{Introduction} \label{sec:intro}

Variable and episodic accretion is recognized as a standard feature of the star formation process \citep{Fischer2022}. This expectation is supported by observations across a broad range of diagnostic signatures, including FU\,Orionis and EX\,Lup type brightening events \citep[e.g.,][]{Herbig1966,Hartmann1996,Audard2014}, water and CO snow lines that extend beyond expectations for present-day luminosities \citep{Lee2007, Jorgensen2015,Hsieh2019}, and multiple clumps within individual jets \citep{Lee2010} and outflows \citep{Plunkett2015}. Furthermore, near-infrared, mid-infrared, and sub-mm monitoring observations provide direct evidence of variable accretion as a common protostar phenomenon \citep{Rice2015,Park21}.

Protostellar jets and outflows are indirect records of the variable accretion history, being natural byproducts of the protostellar accretion process \citep[e.g.,][]{Frank2014,Lee2020,Fischer2022}. Furthermore, the mass ejection rates are seen to be proportional to the mass accretion rates \citep[e.g.,][]{Lee2020}. Within the outflows, high-velocity jets show quasi-episodic knots from which we can derive the timescales of the variability \citep{Plunkett2015,Lee2020,Jhan2022}.

HOPS\,373 is a protostellar binary system located in the Orion B cloud \citep[e.g.][]{Johnstone2001} at a distance of 428~pc \citep{Kounkel2018} showing a modest 30\% brightness variability at 850\,$\mu$m in the James Clerk Maxwell Telescope (JCMT) Transient Monitoring Survey \citep{Yoon2022}. HOPS\,373 consists of two continuum sources with a separation of 3.\arcsec6 (1500\,au at the distance of 428\,pc): referred to as the northeast (NE) and southwest (SW) sources \citep{Tobin2015,Tobin2020,Yoon2022}.
Based on the spectral energy distribution, the estimated bolometric luminosity and the bolometric temperature of the HOPS 373 system are 5.3~$L_\odot$  and 37~K, respectively \citep{Kang2015,Furlan2016}.

Complex outflow features are observed in HOPS\,373. The $^{12}$CO\,2--1 emission shows both a large-scale southeast (SE)-northwest (NW) outflow and a small-scale east-west outflow \citep{Tobin2016,Yoon2022}. The dynamical age and the inclination of the large-scale CO outflow are 10$^4$ years and 40\degree, respectively \citep{Furlan2016,Nagy2020}.  The small-scale CO outflow is associated only with HOPS\,373SW. In contrast, no small-scale molecular emission has been detected in HOPS\,373NE, except for the C$^{17}$O\,3--2 line \citep{Yoon2022,Lee2023}. \citet{Yoon2022} argued that the large-scale SE-NW outflow either is associated with the NE source or that wiggles within the outflow from the SW source changed its outflow direction from a blue-shifted large-scale SE outflow to a red-shifted small-scale east outflow.

High spatial resolution ALMA observations show prominent jets and bow shocks ejected from HOPS\,373SW. In this work, we investigate the properties of these observed jets  and outflows and derive the possible link with the observed variable accretion. The observations and archive data sets are described in Sections~\ref{sec:obs}.  In Sections \ref{sec:jet_morphology}--\ref{sec:property_of_jet}, the properties of SiO and \chxoh\ jets/outflows are investigated. We discuss the stellar mass, episodic accretion, and the origin of the two outflows in Section~\ref{sec:discussion} before summarizing our results in Section~\ref{sec:summary}.

\section{Observations} \label{sec:obs}

We investigated the jets and outflows in HOPS\,373SW using Cycle 7 ALMA observations (2019.1.00386.T, PI: Jeong-Eun Lee) and Cycle 4 ALMA archive data (2015.1.00041.S, PI: John J. Tobin). Detailed observations and data reduction can be found in \citet{Lee2023} for the Cycle 7 data and in previous works \citep{Tobin2020,Yoon2022} for the Cycle 4 data.

The Cycle 7 observations were carried out on 2021 August 2 and 6 with thirty-eight 12-m antennas, and a longest baseline of 5.6~km. The total integration time on the source was 61 minutes. The jet tracers SiO\,8--7 (347.330631~GHz) and \chxohl\ (338.34--338.73~GHz) \citep{Tafalla2010,Tychoniec2021} were covered with this observation. The SiO and \chxoh\ lines were cleaned using the natural weighting with the uvtaper of 0.\arcsec2 ($\sim$455 k$\lambda$) to increase the sensitivity, resulting in the beam size of $\sim$0.\arcsec23 with a rms of $\sim$3~\funit\, over the 1~\kms\, spectral resolution. In addition, two \chxoh\ lines at 338.408698~GHz and 338.344588~GHz were averaged to increase the signal-to-noise ratio (S/N). A high resolution ($\sim$0.\arcsec08) SiO image  with a rms of $\sim$2~\funit\, over the 1~\kms\, was also cleaned using natural weighting  without uvtaper in order to study the jets near the continuum peak.

$^{12}$CO\,3--2 (345.7959899~GHz) was covered by the Cycle 4 archive data, which were observed on 2016 September 3 and 4 and on 2017 July 19. The $^{12}$CO line was cleaned using the natural weighting with the uvtaper of 500 k$\lambda$, resulting in a beam size of  0.\arcsec25 with a noise level of 20~\funit\, over 1~\kms\, spectral resolution.

The Atacama Compact Array (ACA) archive data toward HOPS\,373  were also used to check the extended features. These observations were carried out in March 2019 as a part of the program 2018.1.01565.S (PI: Tom Megeath). Lines of C$^{18}$O\,2--1 and N$_2$D$^+$\,3--2 were covered with a beam size of  7.\arcsec4  $\times$  4.\arcsec9 and a spectral resolution of  0.34 and 0.16 \kms, respectively.

%
%

\section{Morphology of Jets and Outflows} \label{sec:jet_morphology}

\begin{figure*}[t]
\includegraphics[width=0.9\textwidth]{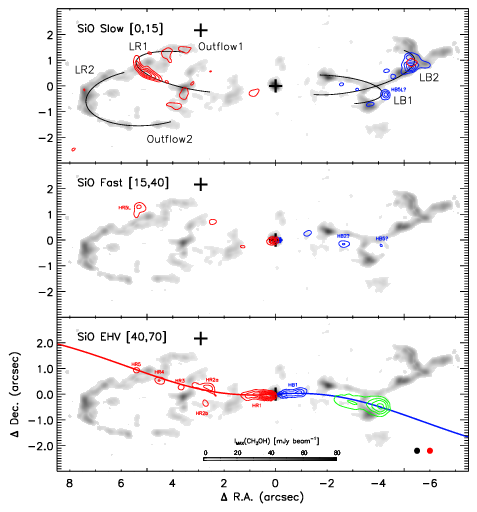}
\caption{ Maps of SiO\,8--7 emission at slow (top), fast (middle), and extremely high velocity (EHV, bottom).
The red and blue contours indicate the red- and blue-shifted SiO emissions integrated over the velocity ranges respective to the systemic velocity presented in each panel. The start and step of contour levels are 10$\sigma$ (1$\sigma$= 13, 17, and 19 \funitv\ for the slow, fast, and EHV components, respectively). 
The background image represents peak intensity (moment 8) map of \chxoh, and \chxoh\ emissions below 5$\sigma$ are clipped out for clarity.
The green contours in the bottom panel indicate the H$_2$ emission \citep{Yoon2022}. The origin and black cross indicate the continuum peak positions of HOPS\,373SW and 373NE, respectively. The locations of the knots are marked in the red (HR1, HR2a, ..., HR5, HR5L) and blue (HB1, HB2, HB5, HB5L) lobes. The black half-ellipses indicate the bow shocks, LB1, LR1, LB2, and LR2.  The beam sizes of \chxoh\ and SiO are presented by black and red ellipses in the lower right corner. The red and blue curves in the bottom panel indicate an example wiggling pattern due to the precession of a putative close binary (see Section~\ref{sec:episodic}).}
\label{fig:jets}
\end{figure*}

\begin{figure*}[t]
\includegraphics[width=1.0\textwidth]{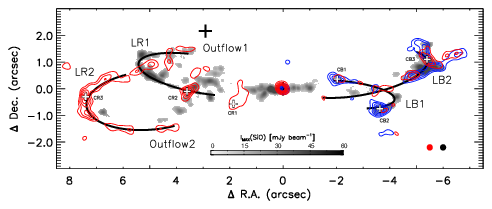}
\caption{ Map of \chxoh\ emission. The red and blue contours indicate the red- and blue-shifted \chxoh\, emissions integrated over the velocity ranges of  0.5 $<|\Delta V|<$3.5 \kms, and the start and step of contour levels are 10$\sigma$ (1$\sigma$= 3.0 \funitv). 
The background image indicates the peak intensity (moment 8) map of SiO 8--7 emission, and the SiO emissions below 5$\sigma$ are clipped out for clarity.
The positions of the strong \chxoh\, emission are marked by white crosses (CR3, CR2, CR1, CB1, CB2, and CB3).  The \chxoh\ emission at the origin is originated by a hot corino as well as the red-shifted outflow \citep{Lee2023}. The beam sizes of SiO and \chxoh\ are presented by black and red ellipses in the lower right corner.} 
\label{fig:jets_chxoh}
\end{figure*}

\begin{figure*}[t]
\includegraphics[width=1.0\textwidth]{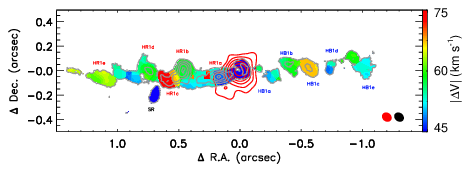}
\caption{ Peak intensity (contour) and absolute velocity at the peak intensity (color) maps for the SiO 8--7 using the high angular resolution image. The maps are created as moments 8 and 9 maps using the CASA task `immoments' after smoothing with a boxcar of 5~\kms. The SiO emission toward the west ($\Delta R.A.< 0\arcsec$) is blue-shifted, but the absolute velocity is plotted for comparison with the red-shifted lobe. The start and step of contour levels are 6$\sigma$ (1$\sigma$= 1.1 \funit). The knots HR1 and HB1 consist of several sub-knots. The emission near the origin is originated from complex organic molecules rather than SiO. The red contours indicate the ALMA Band 7 continuum image. The contours are 10, 20, 40, 80, 160, and 320$\sigma$ (1$\sigma$ = 0.11 \funit). The beam sizes of the continuum and SiO images are plotted in the red and black ellipses in the lower right corner. }
\label{fig:jets_zoom}
\end{figure*}

HOPS\,373SW  shows prominent jets and outflows. Figure~\ref{fig:jets} presents the SiO channel maps over the peak intensity map of \chxoh. In this work, we adopt the systemic velocity ($V_{\rm sys}$) of 10.13 \kms\ derived from the $^{13}$\chxoh\, lines \citep{Lee2023}. All velocities are presented using the relative velocity with respect to this systemic velocity ($\Delta V=V-V_{\rm sys}$). Typically for outflows, the \chxoh\, emission favors a low-velocity gas, while the SiO emission is more abundant in the fastest gas  \citep{Tafalla2010}. For HOPS\,373SW, the \chxoh\, emission has only slow components ($|\Delta V|<$4~\kms)  as shown in Figure~\ref{fig:jets_chxoh}. In contrast, the SiO emission has multiple velocity components: slow ($|\Delta V|<15$~\kms), fast (15~\kms$<|\Delta V|<40$~\kms), and extremely high velocity (EHV, $|\Delta V|>40$~\kms). Note that the observation window covers only up to $-$70 \kms\, from the systemic velocity to the blue, while the red-shifted SiO emission is detected up to $\sim$90 \kms. Thus, we cannot rule out that the highest velocity component in the blue-shifted SiO emission has been missed. Previously, in low mass protostars,  the SiO emission has been observed in clumps \citep{Tychoniec2021} rather than in the continuous outflow denoted by the CO, and these clumps have different peak velocities from the underlying outflow along with narrow line widths ($<$15 \kms). Due to this situation, the distribution of outflow shocks is more clearly revealed by the peak intensity (moment 8 in the CASA task `immoments',  the maximum value of the spectrum in each pixel) maps as shown in Figures~\ref{fig:jets_chxoh} and \ref{fig:jets_zoom}.

\begin{figure*}[t]
\centering
\includegraphics[width=1.0\textwidth]{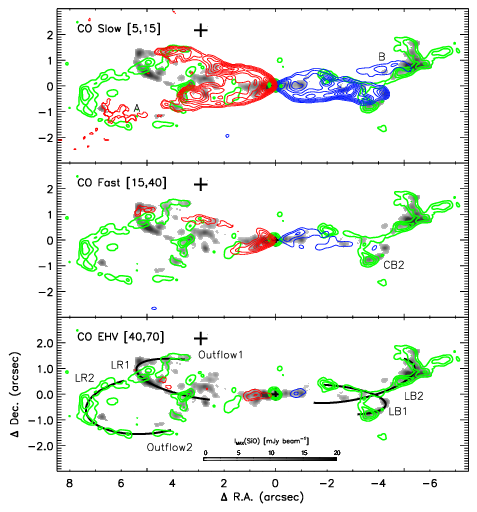}
\vspace{-1cm}
\caption{ Maps of $^{12}$CO\,3--2 emission at slow (top), fast (middle), and extremely high velocity (EHV, bottom).
The red and blue contours indicate the red- and blue-shifted CO emissions integrated over the velocity ranges respective to the systemic velocity presented in each panel. The contours start at 10$\sigma$ with steps of  5 $\sigma$ (1$\sigma$ = 0.04, 0.07, 0.08 Jy~beam$^{-1}$ \kms\, from top to bottom panels, respectively). The background image and green contours indicate the peak intensity (moment 8) maps of the SiO and \chxoh\, emission, respectively,  with 10, 20, and 30 $\sigma$ levels (1$\sigma$= 1.6 \funit). Only the SiO emissions over 5~$\sigma$ are plotted for the clarity. The black crosses indicate the continuum peaks of NE and SW sources, though the contours of the central \chxoh\ peak mostly hide the SW cross. The black curves indicate the U-shaped bow shocks, LR1 and LB1, induced by Outflow1, and LR2 and LB2 by Outflow2. }
\label{fig:co_chan}
\end{figure*}

\begin{figure*}[t]
\centering
\includegraphics[width=1.0\textwidth]{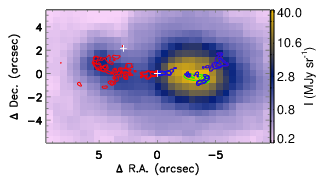}
\caption{Peak intensity map for the red/blue-shifted components of SiO\,8--7 (red and blue contours, respectively) on the Spitzer/IRAC 4.5 $\mu$m image (color). The contours are 5, 10, 20, and 40$\sigma$ (1$\sigma$= 2.4 \funit). The green contours are the Gemini North/GNIRS K-band acquisition image \citep{Yoon2022}, which traces the H$_2$ emission. The white crosses indicate the continuum peaks in the 890 $\mu$m ALMA continuum image.}
\label{fig:spiztermap}
\end{figure*}

The EHV and fast SiO emission components trace the collimated NE-SW jet revealed by the EHV CO emission \citep[see Figure 8 in][ and the bottom panel of Figure~\ref{fig:co_chan}]{Yoon2022}. Figure~\ref{fig:spiztermap} shows that the NE-SW SiO emission is also associated with Spizter/IRAC\,4.5 $\mu$m intensity tracing the H$_2$ emission  in the NIR \citep[green contours  in Figures~\ref{fig:jets} and \ref{fig:spiztermap},][]{Tobin2007,Velusamy2011}. The H$_2$ emission  is detected only in the blue lobe \citep{Yoon2022}. The red-shifted (NE) infrared emission might be extincted by the optically thick protostellar envelope and, therefore much fainter than the blue-shifted (SW) emission.  The  EHV and fast SiO emission consists of several knots in the red (HR1, HR2,...) and blue (HB1, HB2, {and HB5}) lobes, as shown in Figure~\ref{fig:jets}. The EHV SiO knots in the red lobe are well developed  up to a terminal knot, HR5. On the other hand, the EHV SiO knot  HB1 in the blue lobe {is} found only near the protostar, and  the fast SiO knot HB2 and HB5 in the middle panel of Figure~\ref{fig:jets} appear as counterparts to the red-shifted EHV SiO knots HR2 and HR5, respectively. 
The red-shifted SiO jet shows a wiggle from north to south  as reported in the EHV CO map \citep{Yoon2022}, for which we provide a plausible jet precession solution in Section~\ref{sec:episodic}. The SiO and CO emission within the blue lobe shows a point symmetry to those in the red lobe with respect to the continuum peak. 

 Figure~\ref{fig:jets_zoom} shows the peak intensity and the absolute peak velocity maps for HR1 and HB1 near the continuum peak using the high spatial resolution images ($\sim$0\arcsec.08). They are derived after smoothing with a boxcar of 5\,\kms. The knots HR1 and HB1 consist of several sub-knots. In the red lobe, the sub-knot HR1a exhibits a velocity gradient in the southeast direction. HR1b and HR1c appear to be connected, following a curved trajectory. There is an increase in velocity from HR1b to HR1c. HR1d and HR1e are weak compared to HR1a and HR1b. Counterparts in the blue lobe are identified by their distance from the protostar and peak velocity.

 The slow SiO and \chxoh\, emission trace the bow shocks and two distinct outflows as shown in the top panel of Figure~\ref{fig:jets} and \ref{fig:jets_chxoh}, respectively. The NE-SW outflow (hereafter Outflow1) is confined by U-shaped bow shocks, LR1 and LB1 (see the black curves in Figures~\ref{fig:jets}, \ref{fig:jets_chxoh}, and \ref{fig:co_chan}), which could be related to the above jet. The SE-NW outflow (hereafter Outflow2) is also surrounded by U-shaped bow shocks, LR2 and LB2. Similar U-shapes of \chxoh\, and SiO emission are seen in L1157 and they trace the wings of the bow shock \citep{Gomez-Ruiz2013,Codella2020}.

The distribution of SiO and \chxoh\, emission is different between the two HOPS\,373SW outflows.
In Outflow1, the wings of LR1 are mainly traced by the SiO emission, while those of LB1 are traced by the \chxoh\, emission. In contrast, only SiO emission  (e.g., HR5, HR5L, HB5, and HB5L) is detected at the apexes of LR1 and LB1 without corresponding \chxoh\, emission. In Outflow2, the \chxoh\, emission is detected at both the apexes and in the wings. Furthermore, slow SiO emission is clearly detected in LB2 but only marginally in LR2, while the tip of LB2 shows both red- and blue-shifted slow components of  SiO and \chxoh\, emissions implying that the tip might be close to the plane of the sky.

%
%
Figure~\ref{fig:co_chan} shows the CO channel maps overlaid on the SiO and \chxoh\,  peak intensity (moment 8) maps. The EHV CO emission  (bottom panel) traces the jets shown by the EHV SiO emission. The fast (middle panel) and slow (top panel) blue-shifted CO outflow components are confined by the wings of LB1, implying an association with Outflow1. The red-shifted Outflow1 can be traced by the fast red-shifted CO emission and the slow SiO emission (the wings of LR1).  

The CO outflow associated with Outflow2 might be resolved out, except for the observed slow SE (red) and NW (blue) CO outflow components (A and B in the top panel of Figure~\ref{fig:co_chan}).  The velocities may therefore be close to the systematic velocity and Outflow2 oriented close to the plane of the sky, as revealed by the \chxoh\, emission at the tip of LB2. Note that we cannot rule out that all of the slowest CO emission in the map, except for the emission around the apexes of LR1 and LB1, might be related to Outflow2 because the slow CO emission in the north (south) wall within $\sim$3\arcsec\ of the red (blue) lobe seems to connect to the wings of LR2 (LB2). In this case, the outflow velocity might have been decelerated through interaction with the ambient medium and have slowed to close to the systemic velocity near the apex of LR2 and LB2.

\section{Position velocity diagrams of jets and outflows} \label{sec:pv}

\begin{figure*}[t]
\includegraphics[width=1.0\textwidth]{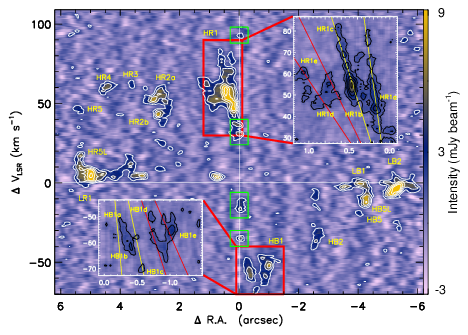}
\caption{Position velocity diagram for SiO emission along the R.A. (jet) direction. Here the emission within $|\Delta$Dec.$|<$1.\arcsec5 is averaged. The start and step of the contour levels are 3$\sigma$ (1$\sigma$= 0.7 \funit). The insert panels are zoomed in the high-velocity jet near the continuum source in the  high angular resolution image. The emission is averaged within $|\Delta$Dec.$|<$0.\arcsec5, and the start and step of the contour levels are 3 $\sigma$ (1$\sigma$= 0.4 \funit).   Both the SiO spectra are smoothed with a boxcar of 3 \kms\ to increase the  S/N.  The emission within the green boxes at the continuum peak are associated with complex organic molecules  \citep[see][]{Lee2023}. The yellow and red lines in the insert panels are the slopes for the ballistic ejection (see the Section~\ref{sec:pv}).} 
\label{fig:pv_ra}
\end{figure*}

Figure~\ref{fig:pv_ra} shows the position velocity diagram along the R.A. (jet) direction within $|\Delta$Dec.$|<$1.\arcsec5. The knots in the red-shifted jet have an observed jet velocity of $\sim$55~\kms\, and a terminal velocity of $\sim$45~\kms\, at HR5. On the other hand, the knots in the blue-shifted jet have decreasing velocities with  $\sim$-55~\kms\, at HB1,  $\sim$-35 \kms\, at HB2, and a terminal velocity of $\sim$-20 \kms\, at  HB5.   Velocity asymmetries in the red and blue lobes are also observed in the SiO jets of L1448C \citep{Yoshida2021} and L1157 \citep{Podio2016}, as well as in the optical/infrared jets from Class II sources \citep{Hirth1994}. This asymmetric velocity could originate from the launching mechanism itself, for example, differences of launching velocity on each side of the disk by the magnetic field \citep{Dyda2015} or the accreting mass \citep{Liu2012}, or as a result of interaction with an asymmetric environment \citep{Podio2016}.  In addition, this could originate due to different inclination angles in the blue and red lobes due to the wiggling of jet. However, the distance of terminal knot HB5 is closer to the protostar compared to HR5; thus, the velocity of the blue-shifted jet should be slower than that of the red-shifted one.

The HR1 knots could have been ballistically ejected. The yellow lines in the upper insert of Figure~\ref{fig:pv_ra} indicate velocity gradients along the distance from the origin with 270\,\kms\,arcsec$^{-1}$ and 120\,\kms\,arcsec$^{-1}$. Therefore the sub-knots HR1a and HR1b would have been ejected 7.5 tan($i$) and 17\,tan($i$) years ago, respectively, where $i$ is the inclination of jet from the plane of sky. The red lines indicate the velocity gradients for ejection events that occurred previously with a similar time interval (9.4\,tan($i$)  years). It is plausible that sub-knots HR1d and HR1e might have been ballistically ejected with the same cadence. On the other hand, sub-knots in HB1 do not show the velocity gradient along the distance despite their proximity to the velocity gradient associated with each counterpart.  The absence of the velocity gradient in the HB1 could be attributed to the limitations in velocity coverage and/or the weaker emission in the blue lobe compared to the red lobe.

\begin{figure*}[t]
\centering
\includegraphics[width=1.0\textwidth]{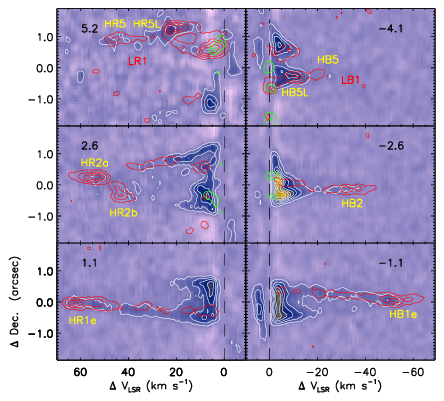}
\caption{Position velocity diagram of  CO (white), SiO (red), and \chxoh\ (green)  emission along the Dec.  direction. The relative distance to the protostar ($\Delta$R.A.) is indicated in each panels. The image and white contours indicate the CO 2-1 emission. The red and green contours indicate the SiO and \chxoh\, emissions, respectively. The start and step of the contour levels are 3 $\sigma$ (1$\sigma$ = 11, 1.7, 1.6 \funit\, for CO, SiO, and \chxoh, respectively). The vertical dashed lines indicate the source velocity. The images are smoothed with a boxcar of 3 \kms.}
\label{fig:pv_dec}
\end{figure*}

Figure~\ref{fig:pv_dec} shows the position velocity diagram of CO\,3--2 (image and white contours), SiO\,8--7 (red contours), and \chxoh\, (green contours) along the Declination (perpendicular to the jet axis).  Near the protostar at $\pm$1.\arcsec1 (bottom panels), both CO and SiO emission connect the high-velocity components (HR1e and HB1e) to the ambient envelope material, implying that the ambient material is being accelerated by the jet \citep{Rabenanahary2022} as also shown in HH 211 \citep{Jhan2021} and Cep E \citep{Schutzer2022}. 
The emission in the velocity range of 20--50 \kms\ is seen only in the southern region in the red-shifted lobe  while no emission is seen in the northern region. The fast CO emission (the middle panel of Figure~\ref{fig:co_chan}) also shows that the red-shifted fast CO emission is extended along southern part while  stunted along the northern part near the protostar. The blue-lobe shows an opposite trend. Furthermore, the red-shifted EHV SiO knots reveal a wiggling pattern from north to south as shown in the red curve in the bottom panel of Figure~\ref{fig:jets}. Therefore, it is plausible that the environment toward the northern part of the red-shifted flow was previously entrained by the earlier north-east jet and outflow, producing the observed asymmetric emission.

Around the HB2 knot (-2.\arcsec6, middle right panel of Figure~\ref{fig:pv_dec}), the SiO emission is similar to that in HB1e, but with smaller jet velocity. The north ($\Delta$Dec.= $+$0.\arcsec3) and south ($\Delta$Dec.= -0.\arcsec4) \chxoh\, emission (green contours) may trace the wings of the bow shocks LB1 and LB2, respectively,  with velocities similar to the systemic velocity. Around HR2 ($+$2.\arcsec6), the knot HR2a might accelerate the north wall traced by the fast SiO and CO emission (see the fast CO outflows in the middle panel of Figure~\ref{fig:co_chan}). The HR2b knot may be interpreted as a southern accelerated component. It is also possible, however, for HR2b to be associated with Outflow2 because it is on the trajectory of the bow shock LR2, as shown in Figure~\ref{fig:jets}.

The southern SiO component indicates the apex of LB1(-4.\arcsec1, top right panel of Figure~\ref{fig:pv_dec}), and is surrounded by two \chxoh\, components at the systemic velocity. Northern SiO emission appears to be associated with LB2. Around the apex of LR1 (+5.\arcsec2, top left panel), we also see the terminal knot HR5 and entrained components. Weak SiO knots toward the south are associated with LR2.

%
%
\section{Jet Models} \label{sec:jet_model}
A simple jet model provides an estimate of the jet physical parameters \citep[e.g.,][]{Raga1990,Lee2004,Jhan2022}. The regular spaced knots can be produced by a periodic velocity variation in which fast-jet material catches up with slow-jet material. This is in agreement with the position velocity diagram which shows higher velocities toward the central source as shown in HR2a, HR4, HB2, then  HB5L in Figure~\ref{fig:pv_ra}.

A time dependent periodic jet velocity $V(t)$ with a fixed period ($P$), mean jet velocity ($V_j$), and velocity variation amplitude($\Delta V_j$) can be written as:
\begin{equation}
    V = V_j - \frac{\Delta V_j}{2}\sin \left(\frac{2\pi t}{P}\right).
\end{equation}
With this model, the distance ($D$) to the first knot  is
\begin{equation}
    D = V_j\frac{V_j}{\Delta V_j/(P/\pi)}
\end{equation}
and the interval ($\Delta D$) between the knots  is
\begin{equation}
    \Delta D = P V_j.
\end{equation}
The observed distance to the first knot ($D_{\rm obs}$) and the observed interval of the knots ($\Delta D_{\rm obs}$) are
\begin{eqnarray}
    D_{\rm obs} &=& D \cos i, \\
  \Delta D_{\rm obs} &=& \Delta D \cos i,
\end{eqnarray}
respectively, where $i$ is the inclination angle of the jet to the plane of the sky.

The observed jet velocity ($V_{\rm obs}$) and the observed sideways-ejection velocity ($\Delta V_{\rm obs}$) are
\begin{eqnarray}
  V_{\rm obs} &=& V_j \sin i, \\
  \Delta V_{\rm obs} &=& \Delta V_j \cos i,
\end{eqnarray}
respectively, when no energy dissipation is assumed and  the sideways-ejection velocity is equal to $\Delta V_j$. 
 Therefore, the parameters of the jets ($V_j$, $\Delta V_j$, $P$, and $i$) can be derived from the observed parameters ($V_{\rm obs}$, $\Delta V_{\rm obs}$, $D_{\rm obs}$, and $\Delta D_{\rm obs}$). For example, the inclination is derived as
 \begin{equation}
     i = \tan^{-1}\left(
     \frac{V_{\rm obs}\,\Delta D_{\rm obs}}{\pi\,\Delta V_{\rm obs}\,D_{\rm obs}}  \right).
 \end{equation}
 
\begin{deluxetable*}{cccc}
\tablecaption{Parameters for the SiO Jet model\label{tb:jetmodel}}
\tablehead{
\multicolumn2c{Measured parameters} & \multicolumn2c{Derived parameters} }
\startdata
$D_{\rm obs}$ &  2.\arcsec8 $\pm$ 0.\arcsec3 & $i$ & 37 $\pm$ 11 \degree \\
$\Delta D_{\rm obs}$ & 1.\arcsec8 $\pm$ 0.\arcsec3 & $V_j$ & 91 $\pm$ 22 \kms \\
$V_{\rm obs}$ & 55 $\pm$ 5 \kms & $\Delta V_j$  & 19 $\pm$ 3 \kms \\
$\Delta V_{\rm obs}$ & 15 \kms  & P & 50 $\pm$ 26 years 
\enddata
 \tablecomments{see Section~\ref{sec:jet_model}.}
\end{deluxetable*}

The jet model is applied to the red-shifted EHV SiO knots HR2a and HR4. Both knots exhibit the velocity pattern  predicted by the jet model, with the higher velocity in the upstream side and lower velocity in the downstream side, as previously mentioned. The knot HR3 is ignored because it does not show such a characteristic velocity pattern. In addition, the intensity of HR3 is lower than that of HR2a and HR4. Note that the knot HR1d seems to have the velocity pattern predicted by the jet model similar to HR2a and HR4, and  might therefore be considered as  a potential candidate for the first knot. However, the distance to the HR1d ($\sim$0.\arcsec8) is shorter than $\Delta D_{\rm obs}$ (1.\arcsec8) while the $D_{\rm obs}$ is generally longer than $\Delta D_{\rm obs}$ \citep{Jhan2022}. In addition, the knot HR1 could be formed by the periodic ballistic ejection as mentioned in Section~\ref{sec:pv}. 
Therefore, the knot HR1d is not used as the first knot in our jet model. 

The observed parameters provide the jet parameters as shown in Table~\ref{tb:jetmodel}. Thus, knots HR2a and HR4 appear to have been ejected $\sim$80 and $\sim$130 years ago, respectively. If HR1a (0.20\arcsec), HR1b (0.48\arcsec), and HR5 (5.5\arcsec) have similar inclination and jet velocity, they were ejected $\sim$6, $\sim$13, and $\sim$150 years ago, respectively.  The former two time scales are consistent with the dynamical time scales of HR1a (5.7 years) and HR1b (13 years) derived from the ballistic ejection in Section~\ref{sec:pv}. Note that when the HR3 is also considered in the jet model, the inclination (20\degree) and the period (13 years) decrease while the jet velocity (155 \kms) increases. The dynamical time scales of knots are reduced by half compared to the jet model without HR3. Then HR1a was ejected in the minimum phase of flux variation \citep[see Figure 4 of][]{Yoon2022}. Thus, we prefer the jet model without HR3.

The knots HB2 and HB5 could be counterparts of the red-shifted SiO jet. Although their velocity is decreasing toward the terminal knot HB5, the projected distances from the continuum peak  are similar to those of the counterpart. When the inclination and jet velocity in the red lobe are adopted, HB2 and HB5 were ejected 72 and 120 years ago, respectively.

\section{Physical and Chemical properties of Jets/Outflows} \label{sec:property_of_jet}

\subsection{SiO and CO}  \label{sec:sio}

\begin{figure*}[t]
\centering
\includegraphics[width=1.0\textwidth]{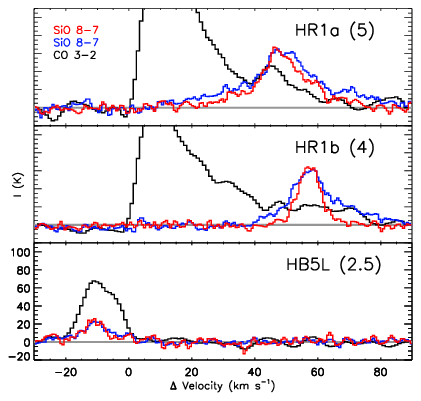}
\caption{Spectra of CO\,3--2 and SiO\,8--7 at knots HR1a (top), HR1b (middle), and HB5L (bottom). The low-resolution (0.\arcsec26) SiO spectra (blue) are the results of convolving the high-resolution (0.\arcsec08) SiO spectra (red) with the same beam as the CO\,3--2 spectra (black) for comparison. The blue and black spectra are multiplied by 5, 4, 2.5 from top to bottom.}
\label{fig:co_sio_spectra}
\end{figure*}

The CO\,3--2 and SiO\,8--7 lines constrain the physical condition within the HR1a knot, which lies 0.\arcsec2 from the continuum peak and appears to have been ejected $\sim$6 years ago. Using the CASA task, `imfit', the high-resolution  SiO 8--7 image (see Figure~\ref{fig:jets_zoom}) shows a spatial width of the HR1a knot of 39 $\pm$ 10 mas (16.7 $\pm$ 4.3 au at the distance of 428 pc), a similar width as that seen in other jets (see Figure 7 in \citealt{Lee2020}). 

We derive the SiO and CO column densities using a Large Velocity Gradient (LVG) model \citep{vanderTak2007,Lee2013}. The radiative and collision coefficients for CO and SiO are provided by the LAMDA database \citep{Schoier2005,Yang2010,Balanca2018}.

 A simple approach tells us that the HR1a knot appears to be dense clump.  The SiO\,8--7 spectrum at the location of HR1a shows a peak intensity of 60 K and a line width (FWHM) of 20 \kms (see the red spectrum in the top panel of Figure~\ref{fig:co_sio_spectra}). Convolved to the beam size of the CO\,3--2 image (0.\arcsec27), the intensity of the SiO emission (blue spectra) is a factor of 5 lower than that of the original (red) SiO spectrum due to beam dilution.  The EHV gas traced by CO\,3--2 and SiO\,8--7 lines show similar spatial distribution in HH211 \citep{Jhan2021}, so we assumed that the CO\,3--2 and SiO\,8--7 lines trace the same gas. The EHV  CO\,3--2 emission from the knot, corrected by the beam dilution factor using the ratio of the red and blue SiO\,8--7 lines, has an  intensity of $\sim$50 K as shown by the black spectrum in the top panel of Figure~\ref{fig:co_sio_spectra}. 
Assuming dense ($\sim$10$^7$ cm$^{-3}$) and warm gas (100--500 K) conditions, the CO column density is 1.1--2.9$\times$10$^{18}$ cm$^{-2}$. Thus, if the CO abundance is 10$^{-4}$ and the length of the line of sight is the same as the knot width (16.7 au), then the derived gas density is 0.4--1.2 $\times$ 10$^8$ cm$^{-3}$, which is consistent with our initial assumption. Further, the mass of HR1a is 0.4--1.1 $\times$10$^{-5} M_\odot$ and the implied mass ejection rate is 1.2--3.2 $\times$10$^{-6} M_\odot$ yr$^{-1}$, when the length of HR1a (123 $\pm$ 13 mas), the inclination (37\degree), and the jet velocity (91 \kms) are adopted. 

The abundance ratio of SiO with respect to CO is different between the EHV and the slow components. 
At HR1a, the EHV SiO column density is 2.7--3.2  $\times$ 10$^{15}$ cm$^{-2}$, and the abundance ratio of SiO to CO is 14--5  $\times$ 10$^{-3}$ (100 K -- 500 K). At HR1b (the middle panel of Figure~\ref{fig:co_sio_spectra}), where the line width is $\sim$10 \kms,  the column density of SiO is 1.3--1.6 $\times$ 10$^{15}$ cm$^{-2}$, and the abundance ratio of SiO to CO is 2--10  $\times$ 10$^{-3}$ (100 K -- 500 K). The EHV knots  (HR2a to HR5)   show a similar line width ($\sim$10 \kms) and a similar intensity ratio between SiO 8--7 and CO 3--2 within a factor of two.

For the slow component (the bottom panel of Figure~\ref{fig:co_sio_spectra}), the abundance ratio decreases by an order of magnitude. The gas density and temperature of the slow component might be between those of the EHV component and those  derived by the excitation analysis of \chxoh\ (see Section~\ref{sec:ch3oh}). 
We applied the same approach used in the EHV SiO component above since the emitting area of the slow SiO component and CO emission at HB5L are similar.
The gas temperature of the slow component is likely higher than 50 K because the peak intensity of the beam-corrected CO is $\sim$ 60 K (see the bottom panel of Figure~\ref{fig:co_sio_spectra}). Thus, we adopt the gas density of 10$^7$~cm$^{-3}$ and the same gas temperature of 100--500 K as in HR1a.  With this setup, the SiO column density is 2.8--4.8 $\times$ 10$^{14}$ cm$^{-2}$ and the abundance ratio of SiO to CO is 1.3--4.6  $\times$ 10$^{-4}$. We note that the red-shifted slow SiO component HR5L is not considered in this analysis because the CO data are too noisy. The EHV component has a higher abundance ratio of SiO to CO than the slow components, which is also seen in Ser-emb\,8\,(N)  \citep{Tychoniec2019}, and the ratios are similar to what is observed in other protostars  \citep{Hirano2010,Podio2021,Dutta2022a,Dutta2022b}.

The SiO could be formed in the EHV gas launched from a dust-poor or dust-free inner disk, where the Si in the dust grains has already been fully liberated into the gas phase, as appears to be the case for other protostellar sources \citep{Lee2020,Podio2021}. In that case, the abundance ratio of SiO to CO of $\sim$5 $\times$ 10$^{-3}$ could be reproduced if the mass ejection rate is $\sim$10$^{-6} M_\odot$ yr$^{-1}$ for the dust-free case or $\sim$10$^{-7}M_\odot$ yr$^{-1}$ for the dust depletion (see Figure 12 in \citealt{Tabone2020}). These ejection rates are consistent with rates estimated above for HOPS\,373SW ($\sim$10$^{-6} M_\odot$ yr$^{-1}$, see the mass ejection rate derived with HR1a).

The abundance of SiO in the slow-moving gas is a factor of 10 lower than that in the EHV gas. For the gas density and temperature derived from the excitation condition, it takes $>$3000 years for the liberated SiO abundance to decrease by a factor of 10 \citep{Simone2022} due to freeze-out onto grain, which is much longer than the dynamical time scale ($\sim$100 years). Thus, the slow-moving SiO component is unlikely to have been originally produced within the jet, where the SiO abundance was high. Instead, the observed slow-moving SiO likely originated from grain sputtering within the bow shock.  The grain sputtering in a C-shock could produce the slow-moving SiO emission  \citep{Gusdorf2008}. This low SiO abundance could be produced within a few tens years according to C-shock models \citep{Gusdorf2008}.

\subsection{CH$_3$OH} \label{sec:ch3oh}

\begin{figure*}[t]
\centering
\includegraphics[width=1.0\textwidth]{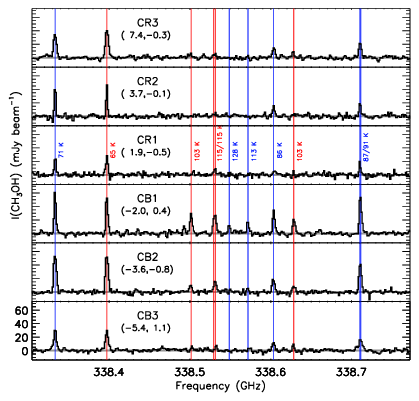}
\caption{\chxoh\, spectra for named positions (white crosses in Figure~\ref{fig:jets_chxoh}) relative to the local continuum peak. The red and blue vertical lines indicate the frequencies of the A- and E- type CH$_3$OH, respectively. The upper energy levels are presented near the vertical lines.  }
\label{fig:spec}
\end{figure*}

Figure~\ref{fig:spec} shows the \chxoh\, 7$_k$--6$_k$ spectra at the \chxoh\,  emission peaks (CR1-3 and CB1-3) presented as white crosses in Figure~\ref{fig:jets_chxoh}. The multiple \chxoh\, lines are fitted by the Large Velocity Gradient (LVG) model \citep{vanderTak2007,Lee2013} using the radiative and collision coefficients for \chxoh\ provided by the LAMDA database \citep{Schoier2005,Rabli2010}. We assume that the abundances of A- and E-type \chxoh\ are the same. The results for the six \chxoh\, peaks are listed in Table~\ref{tb:lvg}. Except for CB1 and CR1, most \chxoh\, emitting regions have a \chxoh\, column density of $\sim$10$^{15}$ cm$^{-2}$, a gas density of 10$^{7}$ cm$^{-3}$, and a gas temperature of $\sim$50 K.

The knots CR1 and CB1 are likely associated with recent shocks. 
The CR1 knot has the highest gas temperature of 155\,K, and the CB1 knot has highest density of 5 $\times$ 10$^7$ cm$^{-3}$.
The \chxoh\,338.559963 GHz, line with the observed highest upper energy level of 128~K, is detected in only CB1. Furthermore, CR1 and CB1 are near the apex of the high-velocity CO outflow and SiO jet near the protostar as shown in the middle and bottom panels of Figure~\ref{fig:co_chan}. Thus, CR1 and CB1 are expected to be recently heated by shocks.

\begin{deluxetable*}{cccccc}
\tablecaption{ LVG models for \chxoh\, emission in HOPS\,373SW \label{tb:lvg}}
\tablehead{
\colhead{} & \colhead{\bf Position}& \colhead{\bf Emitting area$^{a}$}&\colhead{ \bf N(CH$_3$OH)} & \colhead{\bf n$_{\rm H_2}$} & \colhead{\bf T$_{gas}$}\\
\colhead{} & \colhead{(\arcsec,\arcsec)}& \colhead{}& \colhead{10$^{15}$ cm$^{-2}$}&\colhead{10$^7$ cm$^{-3}$} & \colhead{K} }
\startdata
 CR3 & (+7.4, -0.3)  & 0.5 (0.2) & 6.3 (3.2) & 0.8 (1.1) & 30 (12) \\
 CR2 & (+3.6, -0.1)  & 0.3 (0.1) & 10.0 (6.7) & 0.4 (0.4) & 30 (8) \\
 CR1 &  (+1.9, -0.5) &  0.8 (0.2) & 0.6 (1.3) & 1.0 (0.5) & 155 (49) \\
 CB1 &  (-2.0, +0.4) &  0.4 (0.1) & 6.3 (5.3) & 5.0 (1.4) & 85 (18) \\
 CB2 & (-3.6, -0.8)  &  1.0 (0.2) & 2.0 (2.3) & 1.6 (0.3) & 55 (14) \\
 CB3 & (-5.4, +1.1)  &  0.2 (0.2) & 6.3 (15.6) & 0.6 (0.6) & 40 (20) \\
\enddata
 \tablecomments{ The values in the parentheses are the 1$\sigma$ error.\\
 $^{a}$ The unit of the emitting area is relative scale with respective to the beam size of \chxoh.  }
\end{deluxetable*}

\section{Discussions} \label{sec:discussion}

\subsection{Protostellar mass} \label{sec:stellar mass}
The protostellar mass is an important parameter needed to understand the separation and period of any potential binary. The protostellar mass is derived from the Keplerian rotation profile of the gas emission. A rotation feature is shown in the $^{13}$\chxoh\, \citep{Lee2023}. However, the derived stellar mass ($<$0.001 M$_\odot$) is much lower than the theoretical mass of the hydrostatic core \citep[0.01--0.05 M$_\odot$][]{Larson1969,Penston1969,Masunaga1998,Saigo2006} when the Keplerian motion and the inclination of 37\degree\, are adopted. Thus, the observed velocity features near HOPS\,373SW  might  trace the infalling rotating envelope, which has a velocity profile with a steeper power law index (1/$r$) than that of a Keplerian disk (1/$\sqrt{r}$), and HOPS\,373SW might have a very small, as yet unobserved, disk ($<$ a few au).

The protostellar mass can be roughly estimated using an indirect method. 
Class 0 protostars have  a similar ratio of mass ejection rate to mass accretion rate, $\dot{M}_{j}/\dot{M}_{acc}$, \citep[$\sim$0.19,][]{Nisini2015,Lee2020,Podio2021}. The recent mass ejection rate derived in Section~\ref{sec:sio} is 1.2--3.2 $\times$10$^{-6} M_\odot$ yr$^{-1}$, which results in a mass accretion rate of 0.6--1.7  $\times$10$^{-5} M_\odot$ yr$^{-1}$. 
Taking this value for the mass accretion rate, the stellar mass can be estimated from 
\begin{eqnarray}
  M_{star} &= & 0.035\, M_\odot \times \left(\frac{L_{\rm bol}}{5.3 L_\odot} \right) \left(\frac{R_{star}}{2R_\odot}\right) \nonumber \\
  & & \left(\frac{1.2 \times 10^{-5} M_\odot \mathrm{yr}^{-1}}{\dot{M}_{\mathrm acc}}\right) 
\end{eqnarray}
where the bolometric luminosity of 5.3 $L_\odot$ is adopted \citep{Kang2015,Yoon2022}. The stellar mass is then found to be between 0.025--0.07 $M_\odot$.

The derived stellar mass provides a consistent condition launching the SiO jets. The jet-launching radius ($R_{\rm jet}$) is estimated as \citep{Lee2020},
\begin{equation}
    R_{\rm jet} \sim \left(2\frac{\dot{M}_{acc}}{\dot{M}_{jet}} - 3\right) \frac{GM_{\rm star}}{V_j^2},
\end{equation}
where $G$ is the gravitational constant. The derived $R_{\rm jet}$ is $\sim$0.03 au, which is consistent with that in other jets \citep{Lee2020}. The jet-launching radius is much smaller than the dust sublimation radius ($R_s$) of 0.16 au, implying that the jets are launched from the dust-free region as mentioned in Section~\ref{sec:sio}. The dust sublimation radius is calculated as \citep{Monnier2002}
\begin{equation}
    R_s = 7.5 \sqrt{Q_R}\sqrt{L_{\rm bol}/L_\odot}\left(\frac{T_{\rm sub}}{1500 K}\right)^{-2} R_\odot,
\end{equation}
where $Q_R$ is the ratio of dust absorption efficiency at the color temperature of the incident and re-emitting radiation fields and $T_{\rm sub}$ is the dust sublimation temperature. Here, we adopt Q$_R$= 4 and T$_{\rm sub}$= 1500 K \citep{Dullemond2001}.
Note that the values of $M_{\rm star}$, $R_{\rm jet}$, and $R_s$ could be smaller than the values estimated here if the bolometric luminosity of HOPS\,373SW itself is lower than 5.3\,$L_\odot$, which is the total luminosity of HOPS 373 including the SW and NE sources.

\subsection{Dynamical time scales for the two outflows}
\label{sec:dyn-time}

Outflow1 producing the U-shaped bow shocks LR1 and LB1  is younger than Outflow2 producing LR2 and LB2 (see Figure~\ref{fig:co_chan}). The EHV jets and the H$_2$ emission in the near-infrared are associated with Outflow1 (see Figure~\ref{fig:spiztermap}). The jet model shows the knot HR5 at the tip of LR1 in the end of Outflow1 was ejected 150 years ago (see Section~\ref{sec:jet_model}).

The dynamical time scale of Outflow2 can be derived indirectly. The peak intensity and velocity of the SiO knot at the LB2 tip are $\sim$40 K and -3.5 \kms,  respectively, and those of the marginally detected weak SiO knot at the LR2 tip are $\sim$19 K and 4.5 \kms, respectively. The peak temperature of 19--40 K is produced by a $\sim$30 \kms\, shock  if the pre-shock density is 10$^6$~cm$^{-3}$ \citep{Gusdorf2008}. In addition, in LR2 and LB2 the velocity of the peak SiO emission is slightly faster than that of the peak \chxoh\, emission by 0.6\,\kms\ and 2.4\,\kms, respectively. The \chxoh\, is quickly destroyed through reactions with H atoms after sputtering from the grain surface in the high-velocity shock \citep[$>$25 \kms,][]{Suutarinen2014,Nesterenok2018}. Thus, the SiO and \chxoh\, knots would trace the tip and wings of the bow shocks, respectively, as shown in the bow shock LB1, and they might be overlapping each other due to the slow shock velocity.  Using the above estimated shock velocities the dynamical time scale of the Outflow2 can be estimated. For a shock velocity of 30 \kms, the inclination of the outflows is  derived as 7--9\degree\ from the velocities at the peak intensities of LR2 (4.5 \kms) and LB2 (-3.5 \kms), and the LR2 and LB2 would have been ejected $\sim$500 and $\sim$370 years ago, respectively.

\subsection{Episodic accretions}\label{sec:episodic}

The EHV SiO jet shows two periodic variation timescales. The jet model reveals a period of 50 years with an observed separation of 1.\arcsec9 along the HR2a and HR4 knots. The higher resolution image shows that three sub-components lie within HR1 and HR2a (as shown in  Figure~\ref{fig:jets}) where the separation is 0.\arcsec3, a factor of 6 smaller than the larger scale variation and implying a period of $\sim$8 years when the jet velocity and the inclination are assumed the same as the jet model.

These periodic variations may be related to a periodical perturbation of the accretion in the disk by a close binary companion \citep{Lee2020}.
Adopting a protostellar mass of 0.035 $M_\odot$, the variation with $\sim$8 years and $\sim$50 years could be induced by events at 1.3 au and 4.4 au, respectively, which cannot be resolved by the present observations where the major axis of the continuum emission (FWHM) is 71.1 $\pm$ 2.6 mas (30.4 $\pm$ 1.1 au). L1157 MMS, with a separation of 16 au, is the tightest known binary among the directly observed Class 0 protostars \citep{Tobin2022}. Such small scales required detection via VLA continuum observations. Interestingly, the wiggle in the Class 0 protostellar jet HH 211 implies a binary separation of $\sim$4.6 au \citep{Lee2010}. Theoretical modeling finds that the circumstellar disk is truncated by interaction with a binary such that the disk radius is 0.25--0.5 the binary semi-major axis distance  \citep{Terqeum1999}. Thus, for HOPS 373SW the 1.3 au length scale might be associated with the disk radius while the 4.4 au scale represents the separation of the binary. 

This putative close binary could reproduce the wiggle pattern of the SiO EHV emission shown in Figure~\ref{fig:jets}. When a protostar has a mass of $M_p$, a disk radius of $r_{\rm d}$, and a companion is separated with a distance of $a$ and has a mass of $M_c$, then the ratio of precession period $\tau_p$ to orbital period $\tau_o$ is given by :
\begin{equation}
    \frac{\tau_p}{\tau_o}=\frac{32}{15\cos{\beta}}\left(\frac{a}{r_{\rm d}}\right)^{3/2}\left(1 +\frac{M_c}{M_p}\right)^{1/2}\left(\frac{M_p}{M_c}\right),
\end{equation}
where $\beta$ is the half-openning angle \citep{Terqeum1999,Lee2020}. Thus, $\tau_p$ is $\sim$15 times longer than $\tau_o$ assuming $M_p/M_c$=1. The wiggling pattern is described by the parametric equation using $\beta$, $V_j\cdot\tau_p$ (the spatial period), the phase angle ($\phi$), and the inclination \citep[see Eqs (1)-(3) in ][]{Schutzer2022}.   The wiggle motion is roughly reproduced by the precession model assuming $M_p/M_c$=1 and using the parameters $V_j$, $P = \tau_o$, and $i$ from the jet model (Table ~\ref{tb:jetmodel}), $\beta$= 8\degree, and $\phi$= 50\degree\ as shown in the red and blue curved lines in the bottom panel of Figure~\ref{fig:jets}. 
%
%
\begin{figure*}[t]
\centering
\includegraphics[width=0.9\textwidth]{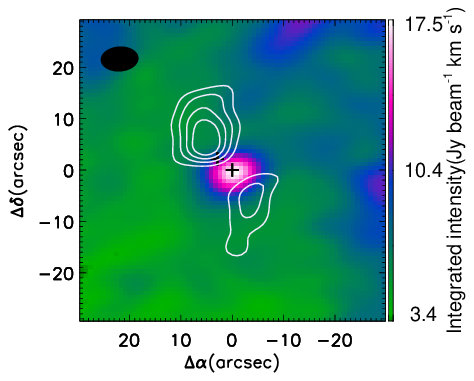}
\caption{The integrated intensity of C$^{18}$O\,2--1 (image) and N$_2$D$^+$ (white contours) observed with the ACA. The C$^{18}$O 2--1 and N$_2$D$^+$ lines are integrated within $|\Delta V|$ $\le$ 5 \kms  and $|\Delta V|$ $\le$ 1.5 \kms, respectively. The contour levels are 8, 12, 16, and 20 $\sigma$ (1 $\sigma$= 0.1 Jy beam$^{-1}$ \kms). The black ellipse indicates the beam size and the black crosses the continuum sources HOPS 373SW and NE.}
\label{fig:aca}
\end{figure*}

According to the relic record provided by the observed knots, many burst events should have occurred in the past. Figure~\ref{fig:aca} shows an anti-correlation between the C$^{18}$O\,2--1 and N$_2$D$^+$ emission.
The C$^{18}$O emission has the Full Width Half Maximum (FWHM) of 4.\arcsec63\,$\pm$\,0.\arcsec61 which is measured by the CASA task, `imfit'. The expected FWHM from the current luminosity (5.3\,$L_\odot$) is 1.\arcsec8 when the CO sublimation temperature is 21\,K \citep{Frimann2017}. Thus, the current CO emission is a factor of 2.6 more extended than that expected from the current luminosity.
In addition, the peak of the N$_2$D$^+$ is 6.\arcsec3--7.\arcsec4 off from the center of HOPS\,373SW, implying the luminosity during a past burst reached $\sim$40--60 \,$L_\odot$ \citep{Hsieh2019,Murillo2022}. 

We note that HOPS\,373SW is very likely the dominant trigger for the extended CO snowline because the distribution of N$_2$D$^+$ emission implies the heating source is closer to the HOPS\,373SW rather than HOPS\,373NE, and the C$^{18}$O 2--1 emission peak is also close to the HOPS\,373SW.   Furthermore, the orientation of the two N$_2$D$^+$  peaks is approximately perpendicular to the large-scale CO outflow in the SE-NW direction, implying that the large-scale CO outflow is also associated with HOPS\,373SW rather than HOPS\,373NE.

The freeze-out time scales for CO and water provide an upper limit for when this large burst occurred. The number density of molecular hydrogen around the observed CO snowline is $\sim$5\,$\times$\,10$^{5}$\,cm$^{-3}$ when the density profile for HOPS\,373 presented in \citet{Furlan2016} is adopted.  At this density, the freeze-out time scale for CO is $\sim$10$^4$ years \citep[e.g.,][]{jelee04}.
On the other hand, the water snowline,  which is derived by the boundary of emission distribution of complex organic molecules ($\sim$0.\arcsec1),  is similar to that derived from the current luminosity \citep{Lee2023}. The water snowline for a large burst (40--60 $L_\odot$) is  $\sim$0.\arcsec23 (100 au at the distance of 428 pc), where the density is higher  compared to that at the CO snowline and the freeze-out time scale for the water is $\sim$100 years \citep{vantHoff2022,Lee2023}. Thus, we expect that this recent burst occurred between the freeze-out time scales for the CO and the water (between $\sim$10$^2$ and $\sim$10$^4$ years). The dynamical time scales of Outflow1 (150 years), Outflow2 ($\sim$500 years), and the large scale CO outflow in the SE-NW direction \citep[$\sim$10$^4$ years,][]{Nagy2020} are all within the estimated time range of the accretion burst that induced the observed CO snowline. Thus, the strongest burst associated with one of the three outflows likely shaped the observed CO snowline. Further observations and studies are needed to determine which outflow is associated with this strongest burst event.

\subsection{Origin of  multiple Outflows}

The SiO, CH$_3$OH, and CO lines toward HOPS\,373SW reveal two outflows: Outflow1 (NE-SW) and Outflow2 (SE-NW).  In addition, the large-scale CO outflow in the  SE-NW direction was also ejected from HOPS\,373SW. A possible explanation for these features is that the three outflows were successively ejected from a single protostar. Initially, the large-scale CO outflow had been ejected $\sim$10$^4$ years ago \citep{Nagy2020}. Then, Outflow2 was ejected  $\sim$370--500 years ago and  lastly Outflow1 was formed  $\sim$150 years ago after the system went through a reorientation, as appears to be the case for IRAS 15398-3359 \citep{Okoda2021}. 
Given that HOPS\,373 is an early stage Class 0 protostar \citep{Kang2015,Furlan2016,Stutz2013}, it expected that the angular momentum axis of the accreting gas might change over time due to in-homogeneities in the star-forming core \citep{Sakai2018, Zhang2019, Gaudel2020, Okoda2021}.
In addition, the outflow reorientation also occurs when the rotation axis of the core is misaligned with the global magnetic field direction \citep[e.g.,][]{Machida2020}. The extended CO snowline suggests that a strong accretion burst occurred 10$^2$--10$^4$ years ago. Therefore, the  three observed misaligned outflows may have been induced by  accretion  bursts accompanied by reorientation. 

A second possible explanation for the two outflows is that they are evidence of  a tight protostellar binary.  In this case as well, the change in the outflow axis from the large scale CO outflow in the  SE-NW direction to Outflow1 or Outflow2 occurred by an accretion burst accompanied by reorientation. The EHV SiO jet shows the wiggling motion, which could be due the precession induced by a close-binary system as mentioned above. Quasi-episodic knots with a period of 50 years also support this hypothesis \citep{Lee2020}. As mentioned above, it is possible that the knot HR2b is associated with Outflow2 and that the \chxoh\ emission peaks CR2, CR1, and CB1 are on the trajectory of the bow shocks LR2 and LB2. If those \chxoh\ peaks are associated with Outflow2, then CR1 and CB1 would have be ejected $\sim$130 years ago, similar to the time scale of the terminal knots ($\sim$150 years) in Outflow1. In addition, the two outflows have different spatial properties. They are misaligned by $\sim$30\degree\, and the jet velocities are $\sim$90 \kms\, and $\sim$30 \kms, respectively. Therefore, each outflow might have been ejected by a different member of a close-in protostellar binary.

Such a misaligned close binary could be formed through core fragmentation. Misaligned outflows with an angle of 70\degree\, have been reported in the Class 0 protostellar binary system VLA 1623 A, where the binary separation is 36 au \citep{Hara2021}. The binary separation in the HOPS\,373SW is estimated to be 4.4 au (see above), which cannot be resolved with the present high-resolution continuum image (FWHM $\sim$34 au). Such a misaligned close binary might be formed initially at larger separations via turbulent fragmentation and have later migrated inward \citep[e.g.,][]{jelee17, Lee2019,Offner2022}.
We note that magnetohydrodynamic simulations \citep{Lee2019,Saiki2020} show that close binaries ($<$ 30 au) can be formed via core fragmentation.
A recent hydrodynamical simulation for a multiple protostellar system also showed that the interaction between a close binary could also result in misaligned disks \citep{jelee23b}.

\section{Summary} \label{sec:summary}

We have investigated the jets and outflows in HOPS\,373SW using high-resolution ALMA observations of the SiO\,8--7, CH$_3$OH, and CO\,3--2 lines. The results are:

(1) The slow moving SiO and \chxoh\, show two distinct bow shocks which surround two separate CO outflows: Outflow1 (NE-SW) and Outflow2 (SE-NW).

(2) The extreme high velocity (EHV) and fast moving SiO emission show several knots, which could be formed in quasi-episodic jets with an inclination of 37\degree, a jet velocity of 91 \kms, and a period of 50 years. These knots also show a wiggle motion previously noted in the CO and H$_2$ emissions by \citet{Yoon2022}.

(3) The SiO jets show two periodic variation timescales with periods of 50 years and 8 years. 
The stellar mass is estimated to be 0.035 $M_\odot$ in which case the longer the periodic variation might be induced by a binary with a separation of 4.4 au. 

(4) The observed extended CO snow line implies that a previous outburst with a luminosity of 40--60 $L_\odot$ occurred between $\sim$10$^2$ and $\sim$10$^4$ years ago. It is possible that the observed CO snowline is  associated with  either the bow shock LR1 and LB1 ($\sim$150 years ago)  or the large scale CO outflow in the SE-NW direction ($\sim$10$^4$ years ago).

(5) The two observed misaligned outflows appear to have either be ejected from a single protostar, due to a reorientation of the angular momentum axis, or from protostellar binary, where each outflow is related to a different binary component. The necessary tight binary would have been formed thorough core fragmentation with a wider separation and after which it migrated inward. 

\section*{Acknowledgements}
This work is supported by the Korea Astronomy and Space Science Institute under the R\&D program (Project No. 2024-1-841-00) supervised by the Ministry of Science and ICT. J.-E. Lee\ is supported by National Research Foundation of Korea (NRF) grant funded by the Korea government (MSIT) (grant number 2021R1A2C1011718). 
G.J.H.\ is supported by general grants 12173003 awarded by the National Science Foundation of China.
D.J.\ is supported by NRC Canada and by an NSERC Discovery Grant. 
The National Radio Astronomy Observatory is a facility of the National Science Foundation operated under a cooperative agreement by Associated Universities, Inc.

This paper makes use of the following ALMA data: ADS/JAO.ALMA\#2019.1.00386.T, ADS/JAO.ALMA\#2015.1.0041.S, and ADS/JAO.ALMA\#2018.1.01565.S.
ALMA is a partnership of ESO (representing its member states), NSF (USA) and NINS (Japan), together with NRC (Canada), NSC and ASIAA (Taiwan), and KASI (Republic of Korea), in cooperation with the Republic of Chile. The Joint ALMA Observatory is operated by ESO, AUI/NRAO and NAOJ.

\bibliography{hops373}{}
\bibliographystyle{aasjournal}



\end{document}